\documentclass[%
twocolumn,
superscriptaddress,
 amsmath,
 amssymb,
 aps,
]{revtex4-1}

\usepackage{ulem}
\usepackage{color}
\usepackage{graphicx}
\usepackage{dcolumn}
\usepackage{float}
\usepackage{bm}
\usepackage[colorlinks = true,
linkcolor = blue,
urlcolor  = blue,
citecolor = blue,
anchorcolor = blue]{hyperref}

\usepackage{color} 
\definecolor{myCmmntColor}{RGB}{173, 0, 147}

\begin{document}


\title{Direct evidence of Klein-antiKlein tunneling of graphitic electrons in a Corbino geometry}

 \author{Mirza M. Elahi}
 \altaffiliation[Present address: ]{Intel Corporation, Santa Clara, CA 95054, USA}
 \altaffiliation[Corresponding author: ]{\href{mailto:me5vp@virginia.edu}{me5vp@virginia.edu}}
  \affiliation{Department of Electrical and Computer Engineering, University of Virginia, Charlottesville, VA 22904, USA}

\author{Yihang Zeng}
\altaffiliation[Present address: ]{Department of Physics, Cornell University, Ithaca, NY, 14850, USA}
\affiliation{Department of Physics, Columbia University, New York, NY 10027, USA}

\author{Cory R. Dean}
\affiliation{Department of Physics, Columbia University, New York, NY 10027, USA}

\author{Avik W. Ghosh}
\affiliation{Department of Electrical and Computer Engineering, University of Virginia, Charlottesville, VA 22904, USA}
\affiliation{Department of Physics, University of Virginia, Charlottesville, VA 22904, USA}
 	
%

\date{\today}
\begin{abstract}
    Transport measurement of electron optics in monolayer graphene p-n junction devices has been traditionally studied with negative refraction and chiral transmission experiments in Hallbar magnetic focusing set-ups. We show direct signatures of Klein (monolayer) and anti-Klein (bilayer) tunneling with a circular `edgeless' Corbino geometry made out of gated graphene p-n junctions. Noticeable in particular is the appearance of angular sweet spots (Brewster angles) in the magnetoconductance data of bilayer graphene, which minimizes head-on transmission, contrary to conventional Fresnel optics or monolayer graphene which shows instead a sharpened collimation of transmission paths. The local maxima on the bilayer magnetoconductance plots migrate to higher fields with increasing doping density. These experimental results are in good agreement with detailed numerical simulations and analytical predictions. 
	%
	\end{abstract}

\maketitle
		\begin{figure*}[t]
    	    \centering
    	    \includegraphics[width=\linewidth]{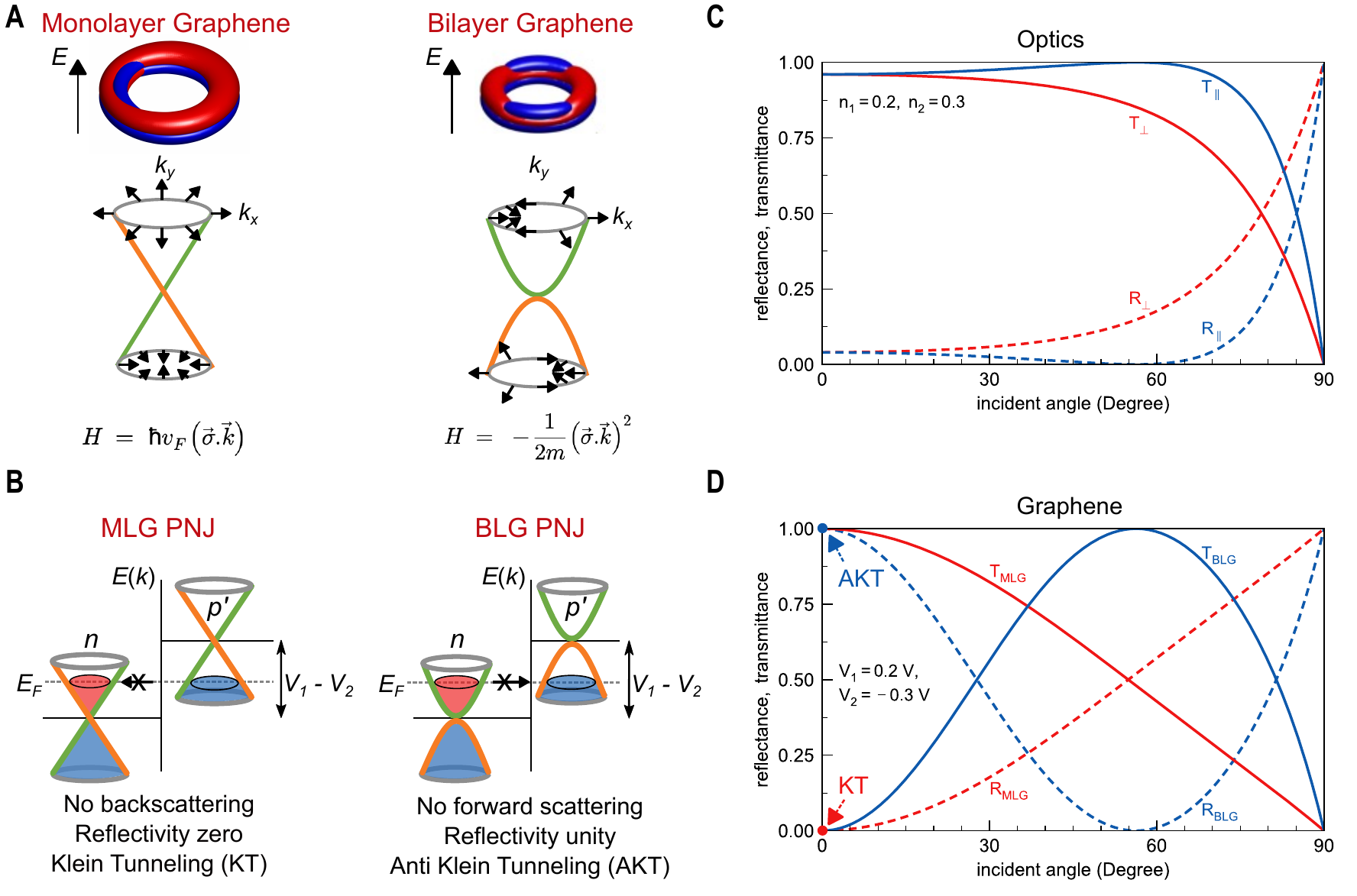}
    	    \caption{\textbf{A schematic for junction transmission in layered graphene showing Klein and anti-Klein tunneling compared with transmission in regular optics.} (\textbf{A}) Pseudospin evolution around the graphene Fermi surface, with red and blue lobes showing positive and negative lobes of dimer p$_z$ orbitals. Bilayer Graphene (BLG) pseudospins rotate twice as fast as in Monolayer Graphene (MLG). Orange vs green branches label left-oriented pseudospin (i.e., antibonding dimer p$_z$ orbitals) vs right (bonding). (\textbf{B}) Schematic of Klein and Anti-Klein tunneling at MLG and BLG split gated p-n junction (PNJ) at normal incidence. (\textbf{C}) Reflectivity $R$ (dashed line) and Transmittivity $T$ (solid) of in-plane vs perpendicular (blue $\parallel$ and red $\perp$) photons across a junction, dictated by Fresnel equations. (\textbf{D}) Electron reflection at a graphene p-n junction (red dashed line in MLG) shows KT at normal incidence, while in BLG (blue dashed line) it shows AKT at normal incidence as well as a Brewster angle set by $\tan^{-1}(V_{G2}/V_{G1}$, here near 60$^0$, where $R_{BLG}$  vanishes. }
    	    \label{fig:ktakt}
            \end{figure*}
        Topological attributes of 2-D Dirac fermions bring to bear a lot of fascinating electronic material properties that are not readily realizable in their optoelectronic counterparts. For instance, the photon-like Dirac cone electron band structure in graphene causes its electrons to bend with spatially varying electron density, much like light bends through a prism or a lens according to Snell's law. However since charges come with dual polarity, the flipping of effective mass around an electrostatically split-gated p-n junction amounts to a flipped sign of the refractive index i.e., the longitudinal wavevector component, suggesting a Veselago focusing even without a lens \cite{cheianov2007focusing}. Realizing such a negative index in optics is much more involved, requiring for instance metamaterials constructed out of split ring resonators operating near absorption thresholds. Furthermore, the transmittivity of electrons is fundamentally different from photons. A change in refractive index causes photons to speed up or slow down, creating an angle-dependent reflectivity quantified by the Fresnel equations.  In contrast, the Dirac cone angle for electrons is set by the strength of the carbon-carbon $p_z$ orbital hopping that does not change at a gated interface, so that electrons can redirect their path but are unable to alter their group velocity in order to reflect backward \cite{cheianov2006selective}. 
       
       The nontriviality of electron topology manifests itself primarily at normal incidence to a p-n junction interface. The Bloch part of the electron or hole wavefunction $u \propto \left(\begin{array}{c} 1 \\ \pm e^{iN\theta}\end{array}
       \right)$  for an $N$-layer Bernally stacked graphene generates a pseudospin structure in the dimer basis set of its frontier carbon p$_z$ orbitals, executing $N$ full rotations around the Fermi surface with a fixed Berry phase $N\pi$ and Chern number of $N/2$. As we will see shortly, this implies the reflectivity at small incident angle $\theta_i$ resembles $\sin{N\theta_i}$ for odd $N$ and $\cos{N\theta_i}$ for even $N$, giving alternately perfect head-on transmission (Klein tunneling or KT) or perfect head-on reflection (anti-Klein tunneling or AKT). This is in sharp contrast to their optical counterpart whose small angle reflectivity is set by the fractional change in refractive index $|n_1-n_2|^2/|n_1+n_2|^2$. The perfect transmission/reflection only works at normal incidence, so that higher angle electrons collimate with increasing voltage barrier (refractive index mismatch) across the split-gated junction. By placing multiple junctions at an angle, the overall transmission can be quenched much like Malus' law in a polarizer-analyzer pair. The result is a gate tunable transmission gap that can be used to turn graphitic electrons off, both for digital \cite{sajjad2013, wang2019} and analog electronic applications \cite{tan2017}. 
       
       While the electronic analogues of Snell's law including negative index \cite{chen2016electron}, and Malus' law \cite{sutar2012} have already been seen, experiments showing transmission peaks are seen only indirectly, through a slight enhancement of transmission in a waveguide geometry \cite{GilHoLee2015}, a predicted increase in resistance of a pair of angled junctions \cite{chen2016electron,wang2019}, an overall peak transmission around normal incidence in  hall bar magnetoresistance data \cite{huard2009}, an anomalous broadening of Fabry Perot peaks in graphene quantum dots \cite{gutierrez2016} and perfect Andreev reflection \cite{lee2019}. A direct demonstration of KT has been missing. Even more striking would be a demonstration of AKT, which would show a highly counter-intuitive absence of transmission at normal incidence unlike any optical analogue, and a maximum transmission at an intermediate gate tunable angle of incidence precisely analogous to the occurrence of Brewster angles in perpendicular polarized light at an interface.  
        
        The direct observation of KT/AKT would ultimately require an edgeless structure such as a Corbino geometry with an out of plane magnetic field to generate clean electron cyclotron orbits entering radially from the center, as opposed to hallbar geometries where electrons enter at the sides and thus face strong restrictions on their incident angles \cite{chen2016electron}. In recent years, bulk properties of graphene have been explored using Corbino geometry, particularly for their ability to resolve fractional quantum hall edge states \cite{schmidt2015second, yan2010charge, zhao2012magnetoresistance, peters2014scaling, kumar2018unconventional}. Corbino structures can be quite useful for probing electron optics in monolayer as well as bilayer graphene due to their high mobility and thus their low scattering probability. 
        
        %
        In this paper, we study magnetotransport in edgeless Corbino devices that include concentric electrostatic p-n junctions at fixed radial distance between the inner and outer electrode.  We measure both monolayer graphene (MLG) and bilayer graphene (BLG) devices, and compare with theoretical calculation and numerical simulations. For MLG, we observe a peaked magnetoconductance versus magnetic field with a clear suppression of transmission in a n-p'-n junction compared to n-n'-n. Strikingly for BLG, we see a dip in some of the magnetoconductance traces around zero magnetic field, and a flattened plateau associated with incipient dips in the other traces. The overall evolution of the magnetoconductance peaks, shoulders and plateau edges vs magnetic fields and doping densities in both sets of data agree very well with theoretical simulations, and serves potentially as a direct evidence of KT and AKT in graphene.

		{\bf{Theoretical background - Klein (antiKlein) tunneling in Mono (Bi) layer graphene Corbino.}}
			The low energy electronic bandstructure for an N-layer Bernally stacked graphene is given by 
		\begin{equation}
            H = \displaystyle\left(\frac{\hbar v_F}{\gamma}\right)^N\left(\begin{array}{cc}0 & (\pi^-)^N \\ (\pi^+)^N & 0\end{array}\right)
        \end{equation}
        with $\pi^{\pm} = (k_x \pm ik_y)/|k| = e^{\displaystyle\pm i\theta}$. The solutions are spinor like (`pseudospin') wavefunctions 
        \begin{equation}
            \psi_C = \displaystyle\frac{1}{\sqrt{2}}\left(\begin{array}{c}1 \\ e^{iN\theta}\end{array}\right), ~~~~ \psi_V = \displaystyle\frac{1}{\sqrt{2}}\left(\begin{array}{c}1 \\ (-1)^{N+1}e^{iN\theta}\end{array}\right)
        \end{equation}
        with C and V representing conduction and valence bands respectively. In Fig.~\ref{fig:ktakt}, the orthogonal states $(1,\pm 1)/\sqrt{2}$, i.e., the bonding and anti-bonding combinations of the frontier dimer p$_z$ orbitals (in-plane for MLG, cross-planar for BLG)  are colored as green and orange respectively. We can then work out the reflection and transmission coefficients across a junction with different wave-vectors tuned by local electron density. 
        Matching the two-component pseudospinors at the interface, we get
    
        \begin{eqnarray}
         T &=& 1-R =  \displaystyle\frac{\cos\left(N(\theta_i+\theta_t)\right) - (-1)^N \cos \left(N(\theta_i-\theta_t)\right)}{1 + \cos \left(N(\theta_i+\theta_t)\right)}\nonumber\\
         &=& \begin{cases}
         \displaystyle\frac{2\cos (N\theta_i)\cos(N\theta_t)}{1 + \cos \left( N(\theta_i+\theta_t)\right)} ~~~(N ~odd) \\
         \displaystyle\frac{2\sin(N\theta_i)\sin(N\theta_t)}{1 + \cos\left(N(\theta_i+\theta_t)\right)} ~~~(N ~even)
         \end{cases}
         \label{eqtran}
        \end{eqnarray}
        where the incident and transmitted angles are related by Snell's law, $\sin{\theta_i}/\sin{\theta_t} = V_{G2}/V_{G1}$ set by the local gate induced potentials on the transmitted and incident sides  
         ($\theta_t$ is negative for pn junctions with opposite gate polarity). For $|V_{G1}| > |V_{G2}|$, $\theta_t$ becomes imaginary for incidence beyond a critical angle and the equations change; $\theta_t \rightarrow \theta_{tR}$ inside the trigonometric functions, and $1 \rightarrow \cosh{\theta_{tI}}$ in the denominator, with subscripts R, I denoting real and imaginary parts. We then get total internal reflection, as the large angle incident electrons cannot conserve their transverse quasi-momenta across the junction. Another modification arises when there is a gradual change in potential across a split gate with separation $d$, picking up an extra angle dependent tunnel transmission factor \cite{cheianov2006selective} $\exp{[{\displaystyle -\pi k_{F\parallel} d\sin{\theta_i}\sin{\theta_t}}]}$, where $k_{F\parallel} = k_{Fi}k_{Ft}/(k_{Fi}+k_{Ft})$ is the parallel combination of the Fermi wavevectors on either side of the junction, amounting to summing the electron transit times across the individual gated regions.

        Note that for all odd $N$, we get Klein tunneling (KT) with perfect transmission (zero reflection) at $\theta_i = 0$ (Fig. 1), while for all even $N$ we get anti-Klein tunneling (AKT, perfect reflection and zero transmission at $\theta_i = 0$).
        These universal values are imposed by the topology of the  electronic phase winding around the Fermi energy, and can be seen simply from the sub-band color matching across the junction (orange can only jump to orange and green to green in 1-D, otherwise the pseudospin arrows do not agree).   
        
        Note also that for even $N$ we get $N/2$ transmission maximizing `Brewster angles' that satisfy
        \begin{equation}
            \theta_i + \theta_t = (2p+1)\pi/N, ~~~p = 0, 1, \ldots N/2-1
        \end{equation}
        while for odd $N$ we also get Brewster angles that satisfy
         \begin{equation}
            \theta_i + \theta_t = 2p\pi/N, ~~~p = 0, 1, \ldots (N-1)/2
        \end{equation}
		This means we expect one Brewster angle for BLG and none for MLG. Much like in optics where a right angle between reflected and transmitted wavevectors quenches reflection (a moving dipole cannot radiate parallel to itself), in bilayer graphene the reflection is quenched since the reflected pseudospin becomes orthogonal to the transmitted one at the Brewster angle, preventing a similar back-radiation. The observation of a transmission minimum at normal incidence followed by a local maximum at a higher, gate tunable Brewster angle, would be a direct experimental signature of AKT. 
	

        \begin{figure}[t]
    		\centering
			\includegraphics[width=\columnwidth]{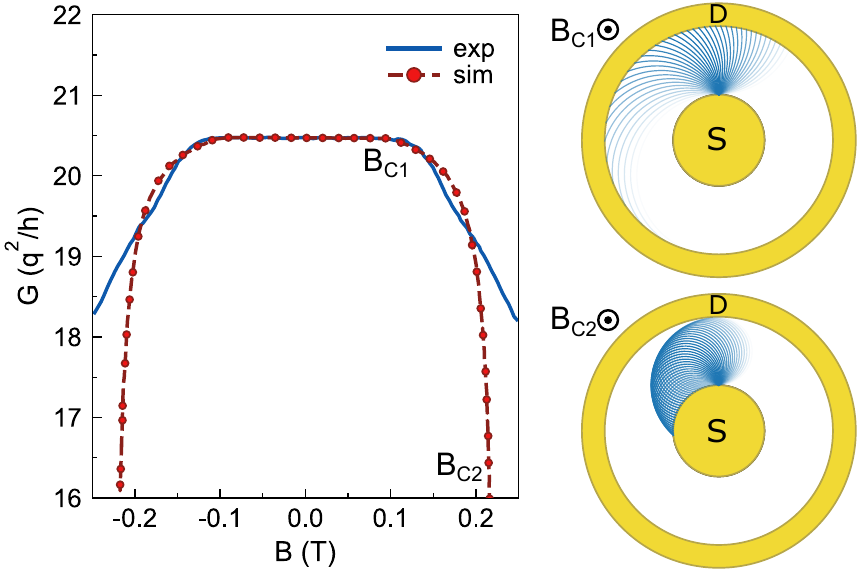}
			\caption{Magnetoconductance plots (solid line experiment, parameters in the text, and dashed line simulation) for junctionless Corbino graphene structure. The cut-off fields $B_{c1}$ and $B_{c2}$ are where the electron cyclotron orbits narrowly miss the furthest and nearest access to the outer contact.}
    			\label{fig:reson}
		\end{figure}

       {\bf{Tracking electrons - Corbino magnetoresonance}} 
       

        Magnetoresistance in ballistic Corbino devices is sensitive to the field modulated electron trajectory. When the tightening cyclotron orbits alternately miss the near and far side contact edges, the conductance drops. For sharp orbits in a perfect Corbino structure, the drops happen at magnetic fields $B_{c1,c2}$ that correspond to that peripheral condition, $B_c = \hbar\sqrt{n\pi}/qr_{c1,c2}$, where the largest tangential orbital radii $r_{c1,c2} = (r_o\pm r_i)/2$, with $r_o$ and $r_i$ being the outer and inner contact radii. We thus expect a big drop in transmission at $B_{c1}$ and a complete quenching of conductance at $B_{c2}$. For Fig.~\ref{fig:reson} shows the response for junctionless corbino device. 
        The experimental data was taken in a junctionless corbino device of monolayer graphene, S1, with $r_i = 1~\mu$m, $r_o = 2.5~\mu$m. Experimental measurement of the magnetoresistance (solid line) shows good agreement with theory (dashed line) exhibiting a broad plateau up to the expected critical field beyond which the cyclotron radius is smaller than the source-drain separation. For fields higher than $B_{c2}$, theory predicts a steep drop in conductance whereas experiment shows a more gradual response. In reality, the electrons enter over a spread of incident angles, the Corbino circles are not perfect (non-concentric and rough around the edges), and the reflected electrons undergo multiple non-specular reflections at the boundaries and  scattering events inside the device. 
        
        The dynamics becomes considerably more complex for Corbino disks with a series of concentric p-n junctions, as needed to see KT/AKT.
        To simulate electron transport for a multi-junction Corbino disk, we adopt a semi-classical numerical ray tracing approach based on a billiard ball model (the arc equation above) that has been used to benchmark experiments in the past \cite{chen2016electron}. Electrons are injected from a circular source (S) at the center of the Corbino disk with a cosine probability distribution in injection angle \cite{molenkamp1990electron}, after which each electron follows classical equations of motion in the gated region, with quantum phases assumed to be suppressed by room temperature decoherence. A full quantum treatment of transport using Non-Equilibrium Green's Functions \cite{Ghoshbook} would be computationally prohibitive \cite{habib2016modified} for micron-sized devices, even with non-atomistic coarse grained grids, especially due to the dominance of incoherent scattering that needs added self-consistency loops. We expect the predominant quantum effects at room temperature in modest B-fields to lie at the junctions, where the topology driven KT/AKT transmission can be written analytically from Eq.~\ref{eqtran}, with exponential modifications for finite junction width \cite{sajjad2012manifestation}. We track each electron, split two-ways into fractional counts by the transmission $T$ and reflection $1-T$ at each junction, and by specularity coefficients at each boundary, until they reach a source or a drain, and extract an overall transmission ${\bar{T}}$ averaged over all incident angles. Using this  transmission and
        number of modes ($M$) we can calculate the low bias channel resistance ($R_{Ch}$) 
		\begin{eqnarray}
		\label{eq:r_akt}
	            M &=& g_v g_s \frac{q(E+V_{ch})}{\pi \hbar v_F} W\nonumber\\
	            R_{Ch} &=& \frac{h}{q^2}\frac{1}{M {\bar{T}}}\nonumber\\
	            R_T &=& R_{Ch} + 2R_C
	            \end{eqnarray}
            where $g_v(=2)$ is the number of valleys, $g_s(=2)$ is the number of spins, $E$ is the energy, $V_{Ch}$ is the channel potential calculated from gate voltage and dielectric thickness, $\hbar$ is the reduced Planck's constant, $v_F$ is the Fermi velocity, $W(=2\pi r_{contact\_in})$ is the circumference of the inner electrode of the device,  ${\bar{T}}$ is the numerically averaged angular transmission per mode, and $R_C$ is the contact resistance extracted from high matched density (n-n-n) bias condition.

        \begin{figure*}[t!]
    		\centering
			\includegraphics[width=\linewidth]{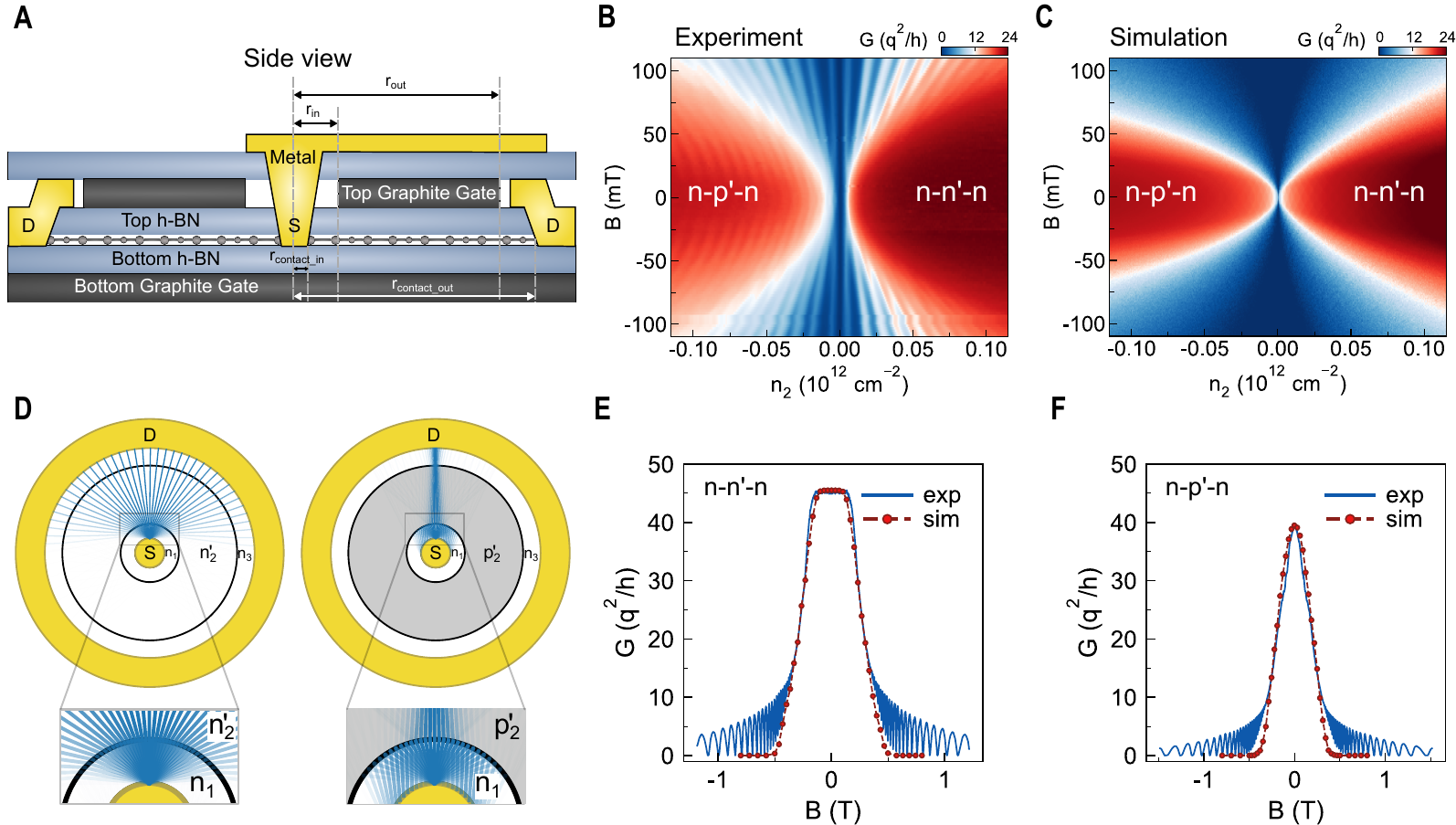}
			\caption[Electron optics in graphene (monolayer) p-n junction with Corbino geometry]{
    			\textbf{Electron optics in Corbino graphene (monolayer) p-n junction.} (\textbf{A}) Side view of the fabricated Corbino monolayer GPNJ device. (\textbf{B}) Experimental and (\textbf{C}) simulated conductance ($G$) colormap between inner and outer contact vs magnetic field $B$ and carrier density  $n_2$. (\textbf{D}) Top view schematic of Corbino monolayer GPNJ device with calculated ray tracing paths (for $B=0$ mT) for n-n'-n and n-p'-n conditions, with zoom plots showing sharply collimated electron transmission for the latter. (\textbf{E}) $G-B$ plot for n-n'-n vs (\textbf{F}) np'n, plotted over a wider range of B-fields at $n_2 = 10^{12}cm^{-2}$. Simulations agree with experiments with two separately benchmarked, i.e., no adjustable parameters (see text). Shubnikov-deHaas oscillations inside a gated region must be separately accounted for in simulations, which focus here on low-bias Klein tunneling at the junction. The results show enhanced transmission with n-n'-n compared to n-p'-n due to Klein tunneling, as seen in (D) from the multiple transmitting paths across the junction. The n-p'-n peak is sharp due to the aggressive collimation from Klein tunneling that restricts transmission at $B = 0$ to normal incidence.}
    	    \label{fig:corbino_monolayer}
			\end{figure*}
			
			{\bf{KT in Corbino MLG devices.}}
            Figure \ref{fig:corbino_monolayer} A shows a schematic cross-section view of the monolayer Corbino device, S2, structure that we measured. The device consists of three circular regions controlled by a bottom and a top gate. The geometry of the top graphite gate is defined into a Corbino shape of inner radius $r_{in}=0.5~\mu$m and outer radius $r_{out}=1.5~\mu$m by electron beam lithography. Similarly the graphene channel is subsequently defined into a slightly larger Corbino with radii $r_{contact\_in}=0.25~\mu$m, $r_{contact\_out}=1.8~\mu$m. Contact resistance $R_C = 256\ \Omega$-$\mu$m. Inner (electron density, $n_1$) and outer ($n_3$) regions are controlled by the bottom gate while the middle region ($n_2$) is controlled by both the top and the bottom gate. In our device, the inner region density $n_1$ is always equal to the outer region density $n_3$ resulting in two symmetric p-n junctions in series between the source and drain. The electrostatic profile of the p-n junction can be controlled by the h-BN thickness, {with screening length $\lambda = \sqrt{(\epsilon_{gr}/2\epsilon_{hBN})t_{gr}t_{hBN}}$} (see methods section for details on fabrication and characterization of Corbino devices).
            
            Figs. \ref{fig:corbino_monolayer}(B, C) show our experimental and simulated magneto-conductance data plotted vs middle region electron density $n_2$ ($n_1 = n_3 = 1.75 \times 10^{12}$ cm$^{-2}$) and magnetic field. Figure \ref{fig:corbino_monolayer}(D) shows our ray tracing simulation paths for electrons for n-n'-n and n-p'-n junctions respectively. For n-p'-n, most of the electrons are filtered out except near zero degree incidences. Figures \ref{fig:corbino_monolayer}(E, F) plot line-cuts  at $n_2 = -1\times10^{12}$ cm$^{-2}$ to show the
            conductance variation versus magnetic field for n-n'-n and n-p'-n cases respectively, along with numerical simulation (semiclassical billiard ray tracing inside each region, angle-averaged quantum transmission following Eq. ~\ref{eqtran} at the interfaces). For n-p'-n, the peak conductance is smaller than n-n'-n due to exponential filtering of electrons for higher angles. The conductance profile is also sharper for the n-p'-n for the same reason. The starting decay point for n-n'-n case can be analytically calculated from Ref. \cite{kirczenow1994quantum}. 
            For both of these conductance simulations, the only two fitting parameters are the split length ($d\sim37$ nm)  of the p-n junction, benchmarked independently from electrostatic simulations, and the two-probe contact resistance that we set at 256 $\Omega$-$\mu$m, also benchmarked from high density experimental resistance. 
            
            For high magnetic fields, our semi-classical results do not match experimental data in the Quantum Hall regime. The Shubnikov-de Haas (SdH) oscillations \cite{tan2011shubnikov} arise from high field quantum interference within the inner gated region, which our semiclassical treatment will miss (our focus is the quantum PN junction tunneling at lower magnetic fields). The high field quantum  resistance is well known, and can easily be reintroduced by adding the two resistances - semi-classical and SdH, in parallel. 
           
        \begin{figure*}[ht!]
    		\centering
            \includegraphics[width=\linewidth]{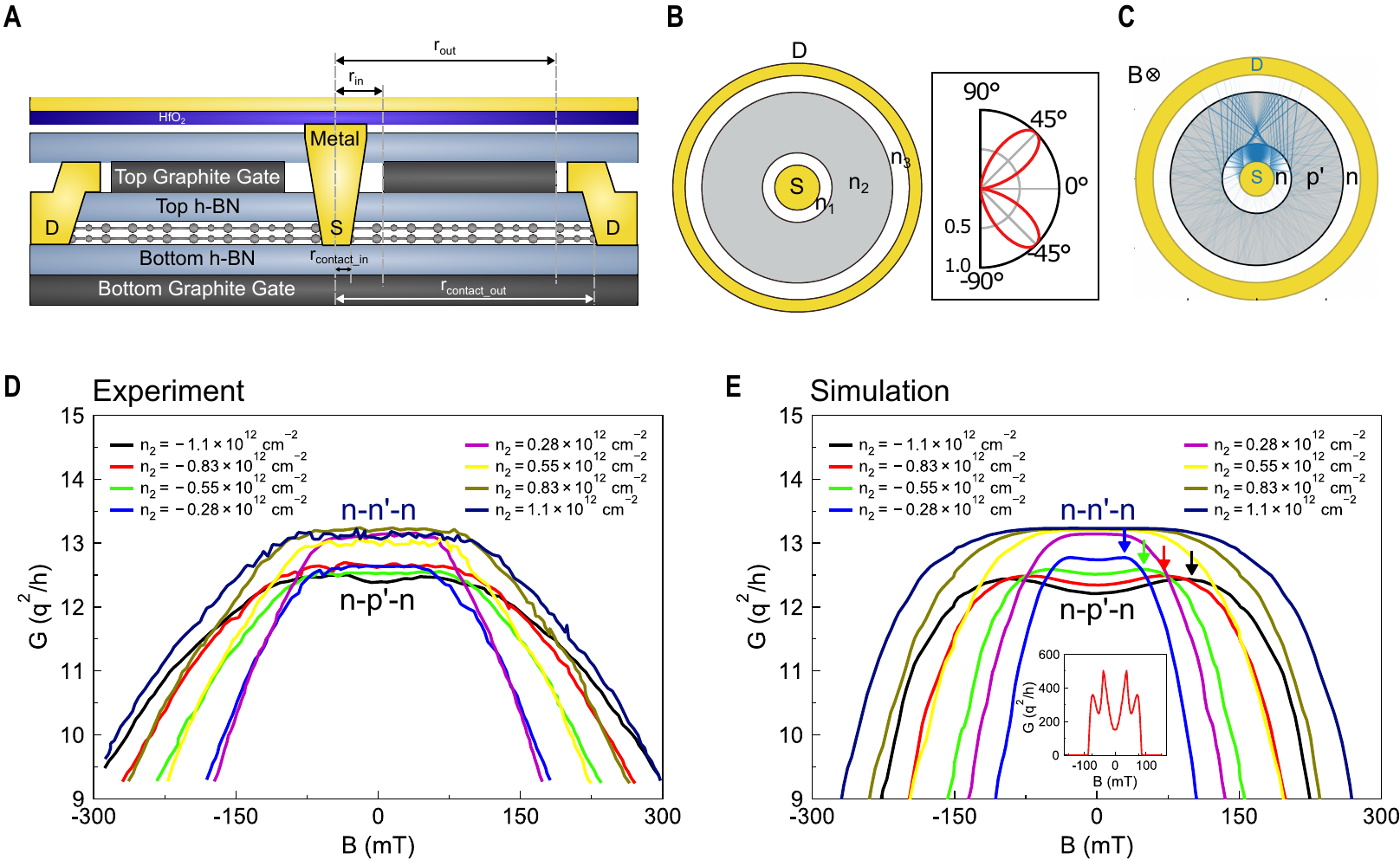}
			\caption{
    			\textbf{Electron optics in Corbino graphene (bilayer) p-n junction.} (\textbf{A}) Side view of fabricated BLG device structure. (\textbf{B}) Top view and polar plot of angle dependent transmission across bilayer graphene p-n junction. (\textbf{C}) Ray tracing paths for n-p'-n device for $B=0$ T. Most of the electrons are filtered across the p-n junction due to anti Klein tunneling, compared to monolayer graphene (Fig.~\ref{fig:corbino_monolayer}D). (\textbf{D}) Experimental vs  (\textbf{E}) simulated $G-B$ plot, simulations including scattering (see text). A pronounced dip is seen at zero field in the magnetoconductance plot corresponding to quenching of the transmission at normal incidence, as expected for BLG AKT (Fig.~\ref{fig:ktakt} blue dashed line). The inset in E shows a simulation with minimal scattering, where the peaks are more pronounced. Arrows show analytical predictions (Eq.~\ref{eq:B_akt}) of the maximizing fields corresponding to the Brewster angle, migrating out to higher fields with increasing doping density.}
    			\label{fig:corbino_bilayer}
			\end{figure*}
			
	{\bf{AKT in Corbino  BLG devices.}} While the above results, in particular the sharpening and shortening of magnetoconductance peaks with added junctions, are in quantitative agreement with our KT simulations in MLG with just two separately benchmarked (i.e., no free) parameters, the qualitative phenomenon - maximum conductance for normal incidence in near-homogeneous junctions, is  expected for any conventional p-n junction. AKT in BLG, however, is completely counter-intuitive to conventional semiconductor heterojunctions, as it predicts maximum transmission at an intermediate, gate tunable Brewster angle.  
	
	Figure \ref{fig:corbino_bilayer}(A) shows a two junction BLG Corbino device, S3, similar to earlier with radii $r_{contact\_in}=0.145~\mu$m , $r_{in}=0.5~\mu$m, $r_{out}=2.5~\mu$m, and $r_{contact\_out}=3.1~\mu$m. 
     Figure \ref{fig:corbino_bilayer}(C, D) show the overall conductance variation versus magnetic field in experiments, compared to simulations (the contact resistance is set at $R_C$=$R_S$=$R_D$=960 $\Omega$-$\mu$m). For n-p'-n, the peak conductance is less than the n-n'-n, both in experiment and simulation, due to the added filtering of electrons at zero angles. This local dip in magnetoconductance at low field is quite unique, and is attributed to anti-Klein tunneling. The inset in Fig. 4E shows the magnetoconductance for a perfectly ballistic structure, zero contact resistance, and electron density set to $n_1 = 2.6\times 10 ^ {12}$ cm$^{-2}$ and $n_2 = -3.2\times 10 ^ {12}$ cm$^{-2}$, while the main plot shows it in presence of scattering (we use an average elastic scattering length of $~$300 nm and a non-specular scattering at each interface extracted from a random Gaussian distribution function with standard deviation $\sigma$ of 20$^\circ$). We would like to emphasize that while the dip itself is seen prominently on one curve, {it happens for parameters where we expect the largest dip, accounting for disorder. Furthermore, we see an abundance of additional curves with robust plateaus that are qualitatively different from Fig. 3 and similar to the simulation results in Fig. 4E, including their overall field and density-dependent evolution. We argue that the evolution of the entire family of magnetoresistance curves and their uncharacteristic shapes including a low-field dip with adjoining shoulders, consistent with numerical simulations and predictable from simple equations (arrow locations in Fig.\ref{fig:corbino_bilayer}E), likely arise directly from anti-Klein tunneling in BLGs.} 
     
            Figure \ref{fig:corbino_bilayer}(B) shows simulated ray tracing paths for electrons for n-p'-n, where most of the electrons are filtered out near normal incidence. 
	        To simplify our calculation for the magnetic field where peak transmission happens in a bilayer graphene p-n junction, corresponding to the Brewster angle, we assume there is no band gap opening due to any significant potential difference between the two layers of graphene (i.e., the vertical fields are screened out). For $\theta_{AKT}=\tan^{-1}(n_1/n_2)$, (where the refractive index $n_i$ is set by the local gate voltage $V_{Gi}$) matches the angle of incidence for which the transmitted and reflected electrons are at angle $\pi/2$, then the conductance is locally maximized. The corresponding magnetic field is given by
	        \begin{equation}\label{eq:B_akt}
	            B_{peak} = \frac{\hbar \sqrt{\pi n_2}}{q r_{AKT}}
	            \end{equation}
	        \begin{equation}\label{eq:r_akt_angle}
	            r_{AKT} = \frac{r_{out}^{2}-r_{in}^{2}}{2r_{out} \cos \left( \frac{\pi}{2} - \theta_{AKT}\right)}
	            \end{equation}
	        In Fig. \ref{fig:corbino_bilayer}(E), we mark out with colored arrows the magnetic field values ($B_{AKT}$) corresponding to the peak conductance points, i.e., Brewster angles, as obtained from Eq. \ref{eq:B_akt}. We see an excellent match between these analytical results and the numerical peak positions, and a good qualitative agreement, especially on the peak migration with doping, compared with experiment (Fig.~\ref{fig:corbino_bilayer}D) and consistent with AKT expectations.\\
\indent{\bf{Conclusion.}} This paper demonstrates direct experimental evidence of Klein and anti-Klein transmission across edgeless (Corbino) PNP mono and bilayer graphene junctions. The physics is expected to get considerably richer with increasing layer number, where Bernally stacked multilayer graphene should show alternate KT and AKT signatures with multiple Brewster angles, albeit over progressively decreasing voltage ranges. Another direction is to look at hybrid structures (mono and bilayer laterally stacked), potentially eliminating low angle electrons with AKT and high angles with KT.   
\\
\indent Nontrivial topological properties of materials have proved critical for their discovery and classification, as well as arguably improving their overall carrier mobility, although these advantages are fundamentally offset by low drivability including carrier density, and poor integrability, including contact resistance. As we point out in this paper, topology also controls the evolution of electron wavefunction around the Fermi surface, which imposes added symmetry restrictions on transmission across barriers. Since barriers can be electrically gated, this opens up the opportunity for pseudospintronics, for unconventional classical computing with topological materials, such as through the deliberate engineering of gate tunable transmission functions. \\
\indent{\bf{Methods.}}
The heterostructure is assembled using the van der Waals transfer technique so that the graphene flake as well as the top and bottom hBN dielectrics are free from contamination during the stacking and subsequent fabrication processes. First plasma etching is used to shape the graphite crystal on top of the heterostructure into a corbino shape. A third hBN crystal is then deposited on top of the heterostructure. Now the final plasma etching is performed to expose the 1D edge of graphene both in the center of the corbino shaped area and around it. Four concentric rings are lithographically defined, the inner and outer radius of graphene edge $r_{contact, in}$, $r_{contact, out}$, the inner and outer radius of top graphite gate  $r_{in}$, $r_{out}$. In monolayer graphene devices, in region with radius $r_{contact, in}<r<r_{in}$ and $r_{out}<r<r_{contact, out}$, the carrier density is defined by the bottom graphite gate $n_{1,3}=C_{bg}V_{bg}/q$. In region with radius $r_{in}<r<r_{out}$, carrier density is defined by both the top and bottom gate $n_2=(C_{bg}V_{bg}+C_{tg}V_{tg})/q$, where $C_{bg,tg}$ stands for capacitance per unit area between bottom, top gate electrodes and graphene respectively and $V_{bg,tg}$ is the voltage bias applied to the top, bottom gate. The monolayer graphene corbino device studied in this work has dimensionality of $r_{in}=0.5~\mu$m, $r_{out}=1.5~\mu$m, $r_{contact\_in}=0.25~\mu$m, $r_{contact\_out}=1.8~\mu$m. The thickness of top hBN and bottom hBN is 31 nm and 116 nm respectively. In bilayer graphene devices, We deposit an additional metal gate on top of a layer of HfO2 dielectric on top of the sample in order to control the electric field in region $r_{contact, in}<r<r_{in}$ and $r_{out}<r<r_{contact, out}$ independent of carrier density, the carrier density is defined by $n_{1,3}=(C_{bg}V_{bg}+C_{cg}V_{cg})/q$ where $C_{cg}$ is the capacitance per unit area between the contact metal gate and graphene and $V_{cg}$ is the voltage bias applied on the metal contact gate. Because the metal gate is completely screened by the top graphite gate in region $r_{in}<r<r_{out}$, the carrier density remains the same as in monolayer samples $n_2=(C_{bg}V_{bg}+C_{tg}V_{tg})/q$. The bilayer graphene device has radii $r_{in}=0.5~\mu$m, $r_{out}=2.5~\mu$m, $r_{contact\_in}=0.145~\mu$m, $r_{contact\_out}=3.1~\mu$m. The thickness of top hBN and bottom hBN is 30 nm and 67 nm respectively. We study the transport properties of the three regions altogether by passing current from the inner contact electrode to the outer electrode as a function of magnetic field. At low temperature ($T < 5$ K), The mean free path in high quality graphene samples can easily exceed 10 $\mu$m \cite{wang2013one}, which is a few times of $r_{contact, out}$ in our sample. Therefore, it is safe to assume quasi-ballistic electron transport behavior at low magnetic field in this work. Since no voltage drop in the sample bulk in ballistic transport, we can extract the contact resistance from the measured two-point resistance when the channel is doped to high carrier density and under zero magnetic field (in deep ballistic regime). The quantum lifetime of electrons can be estimated using the onset of Shubnikov-De haas oscillation at low temperature \cite{Zeng2019corbino}. A typical value in our dual graphite gated graphene device is $0.2-0.3$ ps, corresponding to a disorder energy scale of $13-20$ K. 
\begin{acknowledgments}
    This work was supported by the Semiconductor Research Corporation's NRI Center for Institute for Nanoelectronics Discovery and Exploration (INDEX). Device fabrication and measurement was supported by the Columbia MRSEC on Precision-Assembled Quantum Materials (PAQM) - DMR-2011738.

    \end{acknowledgments}
	
%
  
\end{document}